\documentclass[11.pt]{article}

\usepackage{latexsym}
\usepackage{epsfig}
\usepackage{color}
\usepackage{caption2}
\def\R{ {\rm R \kern -.31cm I \kern .15cm}}
\def\C{ {\rm C \kern -.15cm \vrule width.5pt \kern .12cm}}
\def\Z{ {\rm Z \kern -.27cm \angle \kern .02cm}}
\def\N{ {\rm N \kern -.26cm \vrule width.4pt \kern .10cm}}
\def\1{{\rm 1\mskip-4.5mu l} }
\def\lsim{\raise0.3ex\hbox{$<$\kern-0.75em\raise-1.1ex\hbox{$\sim$}}}
\def\gsim{\raise0.3ex\hbox{$>$\kern-0.75em\raise-1.1ex\hbox{$\sim$}}}
\def\noi{\noindent}

\def\beq{\begin{equation}}   \def\eeq{\end{equation}}
\def\bea{\begin{eqnarray}}  \def\eea{\end{eqnarray}}

\def\noi{\noindent}

\DeclareGraphicsExtensions{.eps}

\def\lsim{\raise0.3ex\hbox{$<$\kern-0.75em\raise-1.1ex\hbox{$\sim$}}}
\def\gsim{\raise0.3ex\hbox{$>$\kern-0.75em\raise-1.1ex\hbox{$\sim$}}}

\begin{document}

\title{\bf Charged multiplicities in pp and AA collisions at LHC}

\vskip 8. truemm
\author{\bf A. Capella$^1$ and E. G. Ferreiro$^2$} 
\vskip 5. truemm

\date{}
\maketitle
 
\begin{center}
\small{
  $^1$ Laboratoire de Physique Th\'eorique\footnote{Unit\'e Mixte de
    Recherche UMR n$^{\circ}$ 8627 - CNRS}, Universit\'e de Paris XI,
  B\^atiment 210, \\
  91405 Orsay Cedex, France
  \par \vskip 3 truemm
   $^2$ Departamento de F{\'\i}sica de Part{\'\i}culas and IGFAE, Universidad de
  Santiago de Compostela, \\
  15782 Santiago de Compostela, Spain}
\end{center}
\vskip 5. truemm

\begin{abstract}
The mid-rapidity charged particle multiplicities in $pp$ and $AA$ collisions at LHC energies are described in the framework of 
a generalized eikonal model with shadowing corrections incorporated in $AA$. We show that the $pp$ data require a Pomeron intercept close to 1.2, higher than the conventional one close to 1.1. An $s^{0.11}$ energy dependence is obtained in the LHC range and beyond. The size and centrality dependence of the $AA$ multiplicity at $\sqrt{s} = 2.76$~TeV is reproduced and its energy dependence is predicted.
\end{abstract}
\vskip 3 truecm

\section{Introduction}
\hspace*{\parindent}
\pagestyle{myheadings}
Recent data on mid-rapidity charged particle multiplicities in proton-proton collisions by the ALICE \cite{1r,2r}, CMS \cite{3r} and ATLAS \cite{:2010ir} collaborations have extended the energy range of this observable up to $\sqrt{s} = 7$~TeV. They show an $s$-dependence in $s^{0.11}$. The same dependence is observed at lower energies. Such an $s$-dependence, observed at high energies, has important consequences for the Pomeron intercept. This is based on a very general property of multiple scattering models, known as the AGK cancellation \cite{4r}. As a result of this cancellation, absorptive corrections (i.e. multiple-scattering effects) vanish identically in the single particle inclusive cross-section $d\sigma/d\eta$. This theorem is valid for a large class of multiple-scattering models, including eikonal and Glauber type models. It is the equivalent in soft processes of the factorization theorem in perturbative QCD. \par

Consider the charged multiplicity $dN^{pp}/d\eta = (d\sigma^{pp}/d\eta)/\sigma_{pp}^{ND}$. The data show an $s^{0.11}$ behaviour for $dN^{pp}/d\eta$. Assuming that the total non-diffractive $pp$ cross-section $\sigma_{pp}^{ND}$ behaves as $s^{\alpha_\sigma}$ with $\alpha_\sigma \sim 0.07-0.09$, as observed at energies below LHC, we conclude that $d\sigma^{pp}/d\eta$ behaves as $s^\Delta$ with $\Delta \sim 0.18-0.20$. Due to the AGK cancellation this is the behaviour of single scattering (Born term) and thus the Pomeron intercept is 
$\alpha_P(0) = 1 + \Delta$
\footnote{A similar pomeron intercept, $\alpha_P(0)=0.2$ \cite{Kai2}, has been prevoiusly used 
by ALICE collaboration \cite{ALICE2} to fit the central rapidity $\bar{p}/p$ ratio as a function of 
the center-of-mass
energy.}. \par

String models like DPM \cite{5r} or QGSM \cite{6r} do, of course, satisfy the AGK cancellation. This cancellation is fully effective at mid-rapidities and high energies. Energy conservation introduces violations in the fragmentation regions which, at low energies, propagate to mid rapidities. It turns out that some amount of violation is still present in mid-rapidity AuAu collisions at RHIC \cite{7r} which go away at mid-rapidities in the LHC energy range.\par

Consider next processes involving nuclei. Here the AGK cancellation implies that the charged multiplicities $dN^{AA}/d\eta$ scale with the number of binary collisions, i.e. the same scaling behaviour valid in hard processes due to the factorization theorem. It is well-known that such a scaling is not supported by the data. Actually, the recent ALICE data \cite{Aamodt:2010cz}, lately confirmed by CMS \cite{Chatrchyan:2011pb} and ATLAS \cite{:2011yr}
collaborations, show that the mid-rapidity charged multiplicity per participant pair $R_A = dN^{AA}/d\eta/(n_{part}/2)$ in central PbPb collisions at $\sqrt{s} = 2.76$~TeV is about two times larger than in $pp$ collisions at the same energy. However, the AGK cancellation implies that this ratio should be equal to the ratio $n_{coll}/(n_{part}/2)$ which is about five at RHIC and increases with energy. The factor of two in $R_A$ relative to $pp$, conveys the idea of a greater efficiency of $AA$ collisions for particle production in each binary collision. However, the surprising fact  is the strong suppression of the $AA$ multiplicity as compared to the predictions of the eikonal-Glauber model. The solution of this puzzle is the presence in $AA$ collisions of the so-called shadowing corrections, not included in that model.
These corrections are small in $pp$ but are quite large in $AA$ collisions where they are enhanced by $A^{1/3}$ factors. In view of that, it is clear that an analysis of the results mentioned above puts strong constraints on the size and energy dependence of shadowing. \par

The plan of the paper is as follows. In Section 2 we describe the physical bases of both AGK cancellation and shadowing. In Sections 3 and 4 we study the mid-rapidity $pp$ and $AA$ charged multiplicities, respectively. Conclusions are given in Section 5.

\section{AGK cancellation and shadowing}
\hspace*{\parindent}
Let us recall the physical bases of the AGK cancellation \cite{4r} and shadowing \cite{7r,9r}.\par

A simple way to illustrate the AGK cancellation is to consider the cross-section $\sigma_\nu^{pA}(b)$ for $\nu$ inelastic collisions in $pA$ scattering in the probabilistic Glauber model
\beq
\label{1e}
\sigma_\nu^{pA}(b) = {A \choose \nu} \left ( \sigma_{pp} \ T_A(b) \right )^\nu \left ( 1 - \sigma_{pp} \ T_A(b)\right )^{A-\nu} \ .
\eeq

\noi This formula is self-explanatory. $T_A(b)$ is the nuclear profile function normalized to unity. The first factor in eq. (\ref{1e}) is the number of ways to choose $\nu$ interacting nucleons out of $A$. The second one in the probability that $\nu$ nucleons interact at fixed impact parameter $b$. The third factor is the probability for no interaction of the remaining $A-\nu$ nucleons. Upon summation of (\ref{1e}) from $\nu = 1$ to $A$ and integration in $b$ one obtains the non-diffractive $pA$ cross-section. It can be seen numerically that it behaves like $A^\alpha$ with $\alpha \sim 2/3$. On the contrary for the multiplicity we have
\beq
\label{2e}
{dN^{pA} \over d\eta} = {1 \over \sigma_{pA}} \ \sum_{\nu = 1}^A \ \nu \ \sigma_\nu^{pA} \ {dN^{pp} \over d\eta} \equiv \overline{\nu}\ {dN^{pp} \over d\eta} 
\eeq

\begin{figure*}[htb!]
\begin{center}
\begin{flushleft}
\includegraphics[width=1.\textwidth]{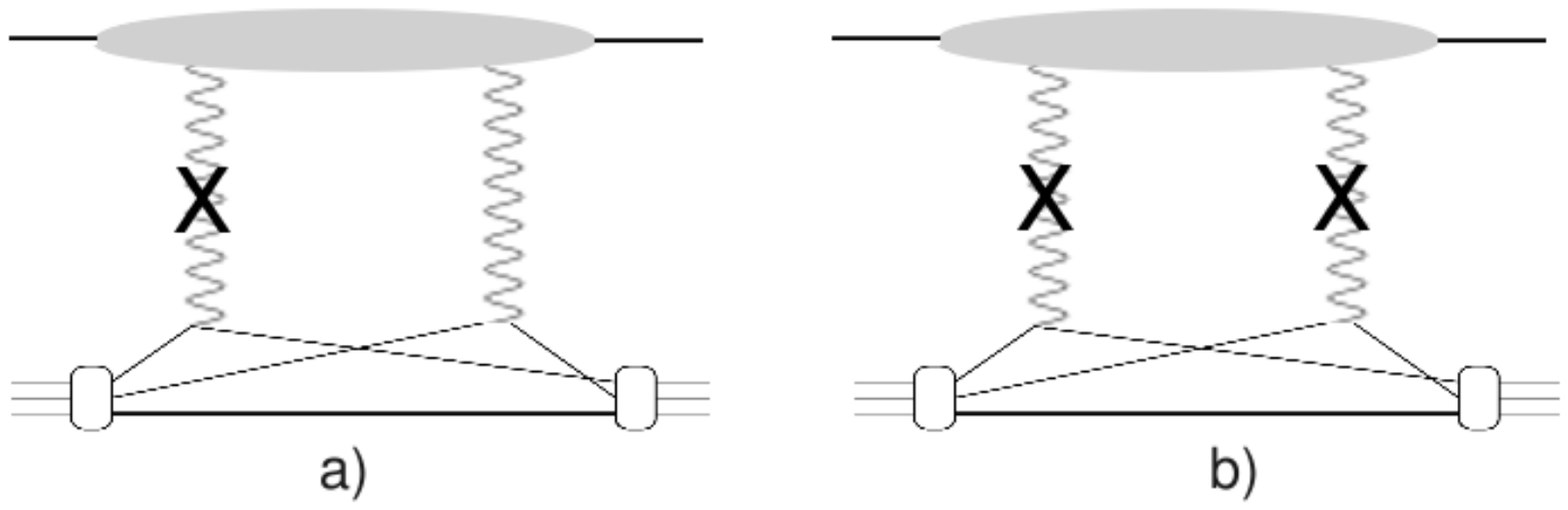}
\end{flushleft}
\end{center}
\vskip -17.75cm
\caption{ 
A double scattering (two-Pomeron exchange graph) in $pA$ collisions with a single inelastic collision (one cut Pomeron), a), and two inelastic collisions (two cut Pomerons), b).}
\label{fig1}
\end{figure*}

\noi with $\overline{\nu} = A \sigma_{pp}/\sigma_{pA}$. The factor $\nu$ in eq. (\ref{2e}) is due to the fact that, in the presence of $\nu$ non-diffractive inelastic collisions, the trigger particle can be produced in any of them. Eq. (\ref{1e}) is equivalent to $d\sigma^{pA}/d\eta = A\ d\sigma^{pp}/d\eta$. Thus $d\sigma^{pA}/d\eta$ scales with $A$ and $dN^{pA}/d\eta$ scales with the number of binary collisions. In the former observable all multiple scattering contributions cancel identically and one is left with the scaling in $A^1$ of the Born term. The same cancellation takes place in $AA$ collisions in the Glauber-Gribov model (see Section 4) and in $pp$ in the eikonal model (see Section 3). In the framework of Reggeon Field Theory \cite{10r} the AGK cancellation is illustrated in Fig.~\ref{fig1} for a double scattering. 
\vskip 0.5cm
\noindent
The first graph, with one cut Pomeron ($\nu = 1$), corresponding to a single inelastic collision, has a weight $-4C$ while the second one, corresponding to two inelastic collisions, has a weight $+2C$. In the summation $\sum\limits_{\nu = 1}^A \sigma_\nu^{pA}$ there is a net contribution $-2C$ (absorptive correction) which reduces the $A$ dependence, while in $\sum\limits_{\nu = 1}^A \nu \ \sigma_\nu^{pA}$ the net contribution is zero. This result is true for any multiple scattering graph and for a very general class of blobs in the upper part of the diagram. This is the origin of the AGK cancellation~\cite{4r} in Reggeon Field Theory~\cite{10r}. \par

\begin{figure*}[htb!]
\begin{center}
\begin{flushleft}
\includegraphics[width=1.\textwidth]{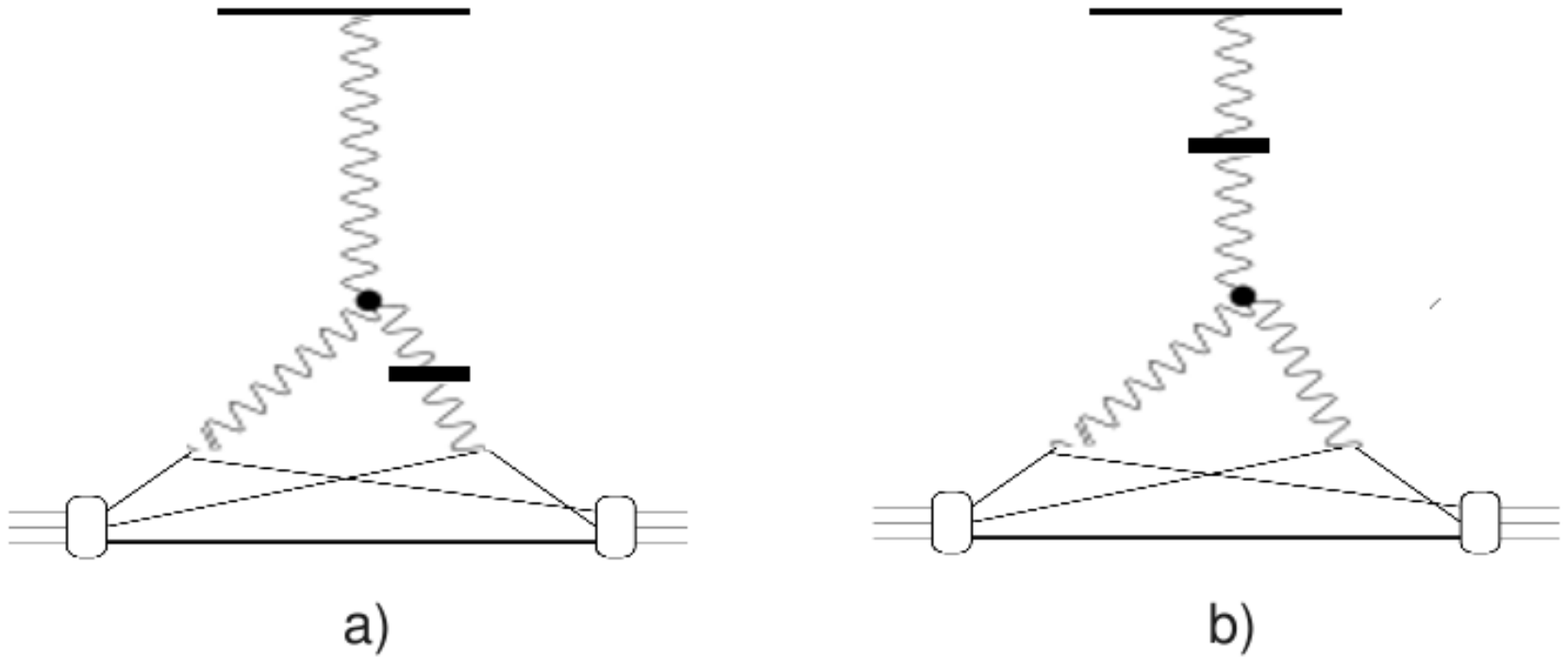}
\end{flushleft}
\end{center}
\vskip -17.75cm
\caption{
A triple Pomeron graph for $d\sigma^{pA}/d\eta$ with the laboratory rapidity of the produced particle smaller (a) or larger (b) than the one of the triple Pomeron vertex.}
\label{fig2}
\end{figure*}

\vskip 0.5cm
Let us turn next to shadowing effects. They are related to graphs associated with large-mass diffraction, commonly known as triple Pomeron graphs, as the ones represented in Fig.~\ref{fig2}. 
In Fig.~\ref{fig2}a the trigger particle is produced at a laboratory rapidity smaller than the one of the junction of the three Pomerons. In this case we are in the same situation as in the graphs of Fig.~\ref{fig1} and the AGK cancellation takes place. The contribution of this type of graphs to $d\sigma^{pA}/d\eta$ cancels identically and we recover an $A^1$ behaviour. On the contrary, in the case of Fig.~\ref{fig2}b, the trigger rapidity is larger than the one of the triple Pomeron vertex and the factor $\nu$ is not present. In this case absorptive corrections are at play which reduce both the size and the $A$ dependence of $\sigma^{pA}/d\eta$. The contribution of the triple Pomeron graphs in Fig.~\ref{fig2} can be computed using parameters determined from large-mass single diffraction data. Following refs. \cite{7r,9r} we use Schwimmer unitarization scheme \cite{11r}. The suppression from shadowing in $AA$ collisions for a particle produced at mid-rapidity ($y^* = 0$) is then obtained replacing the nuclear profile function $T_{AA}(b) = \int d^2s T_A(s) T_A(b-s)$ by
\beq
\label{3e}
S^{sh}(b,s) = \int d^2s {T_A(s) \over 1 + A \ F(y^*=0)T_A(s)} \ {T_A(b-s) \over 1 + A\ F(y^*=0) T_A(b-s)} 
\eeq

\noi where
\beq
\label{4e}
F(y^*=0) = C \left [ \exp (\Delta y_{max}) - \exp (\Delta y_{min}\right ] /\Delta
\eeq

\noi is the triple Pomeron graph contribution, with the rapidity of the triple Pomeron vertex integrated up to $y^* = 0$ (where the trigger particle is produced), i.e. up to $y_{max} = y^* + \ell n (\sqrt{s}/m_T)$. We take $y_{min} = \ell n (R_A m_N/\sqrt{3})$. Smaller values of $y$, corresponding to larger values of the mass of the diffractively produced system, are cut-off by the nuclear form-factor ($t_{min}$ effect). Here $m_T$ is the transverse mass of the particle\footnote{We take the transverse mass of the pion, $m_T=0.38+0.0233*\ell n (\sqrt{s}/53)$ GeV.}, $m_N$ the nuclear mass and $R_A = 0.82~A^{1/3} + 0.58$~fm - the gaussian nuclear radius. $T_A(b)$ is the nuclear profile function for which we use a Woods-Saxon parameterization \cite{12r}. 

Let us discuss now the values of $\Delta$ and $C$. In previous works \cite{9r} we have used $\Delta = 0.13$ and $C = 0.04$~fm$^2$ ($C/\Delta = 0.31$ fm$^2$) corresponding to the Pomeron intercept $\alpha_P (0) = 1.13$. As discussed in the Introduction we show in Section 3 that the LHC data require a higher Pomeron intercept $\alpha_P (0) = 1.19$. Consequently we take $\Delta = 0.19$. Note that we have already used a similar value of $\Delta$ in \cite{13r} to describe $\gamma^*p$ data, $\Delta = 0.2$. With this value of $\Delta$ introduced in the triple Pomeron a good description of the $M^2$-dependence of high-mass diffraction production at $Q^2 = 0$ was obtained\footnote{Note that an $(1/M^2)^{1.1}$ dependence for large-mass diffraction has been obtained \cite{14r} from a fit of the data. However, it has been shown \cite{15r} that the power of $1/M^2$ is larger in the non-absorbed (triple Pomeron) graph than in the absorbed one, i.e. the one measured in experiment.}. This increase in $\Delta$ leads to a decrease in the residue of the Pomeron --to which the constant $C$ is proportional. This induces a reduction of its value to $C = 0.03$~fm$^2$. With these values of $C$ and $\Delta$ the values of the shadowing suppression factor differ from the old ones by less than 10\% in the energy range between $\sqrt{s} = 200$~GeV and 5.5 TeV.\par

We see from eq. (\ref{3e}) that shadowing corrections are enhanced in nuclear collisions by factors $A^{1/3}$, and for this reason, they turn out to be quite important in this case. They are small in $pp$ and will be neglected in the calculations developed in Section~3.

\section{Multiplicities in $pp$ collisions}
\hspace*{\parindent} 
Multiple scattering in hadron-hadron collisions is usually described in a generalized eikonal model in which inelastic intermediate states are included in the vertex functions of the multiple scattering graphs. This corresponds to the inclusion of diffractively produced intermediate states. In Reggeon field theory the large-mass diffractive states correspond to triple Pomeron (see Fig.~2) and loop graphs. A description of this approach, with exponential residues in $t$ --corresponding to gaussians in impact parameter-- can be found for instance in \cite{13r} and in Appendix B of ref. \cite{5r}. A simpler albeit cruder approach is obtained by neglecting the contributions involving triple Pomeron couplings and including low-mass diffractive contributions as intermediate states in the eikonal model \cite{6r,16r}. Using exponential residues in $t$, and a Regge behaviour for single scattering (Born term) in $(s/s_0)^{\alpha_P (t) - 1}$ with the Pomeron intercept $\alpha_P(t) = 1 + \Delta + \alpha '_P t$, all the loop integrals can be performed analytically. The cross-sections $\sigma_k$ for $\kappa$ non-diffractive inelastic collisions --the equivalent of $\sigma_\nu^{pA}$ in eq. (\ref{1e})-- are then given by \cite{6r,16r}
\beq
\label{5e}
\sigma_k (\xi ) = {\sigma_P \over kZ} \left [ 1 - \exp (-Z) \sum_{i=0}^{\kappa - 1} {Z^i \over i !} \right ] \qquad (\kappa \geq 1) \ .
\eeq
\vskip 0.2cm 

\noi Here $\xi = \ell n (s/s_0)$ with $s_0 = 1$~GeV$^2$, $\sigma_P = 8\pi \gamma_p \exp (\Delta \xi )$, and $Z = 2C_E\gamma_p \exp (\Delta \xi ) /(R^2 + \alpha '_P \xi )$.\par

\noi 
From (\ref{5e}) we obtain the non-diffractive inelastic cross-section 
\beq
\label{5eb}
\sigma_{pp}^{ND}(\xi ) = \sum\limits_{\kappa \geq 1} \sigma_\kappa (\xi ) \ . 
\eeq

\noi 
The mid-rapidities $dN^{pp}/d\eta$ is proportional to $\overline{\kappa} = \sum\limits_{\kappa \geq 1} \kappa \ \sigma_\kappa /\sum\limits_{\kappa \geq 1} \sigma_{\kappa}$, as we will explain below.\par

\vskip 0.2cm
An important parameter in (\ref{5e}) is the Pomeron intercept. We take $\Delta = 0.19$. Such a large value has been motivated in the Introduction. We also take $\alpha '_P = 0.25$~GeV$^{-2}$ and $R^2 = 3.3$~GeV$^{-2}$ \cite{6r, 16r}. 
These parameters control the $t$-dependence of the elastic peak and its energy dependence (shrinking).
With these values for $R^2$ and $\alpha '_P$ one obtains \cite{6r} a value of the elastic pic slope $B\simeq 20$ GeV$^{-2}$ at $\sqrt{s}=7$ TeV, in very good agreement with the recent TOTEM measurement \cite{TOTEM}. Note that this value of $B$ is larger than the one in the Born term due to the effect of the unitarity corrections. 
The parameter $\gamma_p = 0.85$~GeV$^{-2}$ controls the size of the interaction cross section
and has been determined 
from the absolute normalization of $\sigma^{ND}_{pp}$, which leads to the values given in Table~\ref{tab1}. 
Note that the energy dependence of $\sigma^{ND}_{pp}$ is essentially 
determined by the Pomeron intercept $\Delta$, and it is about $s^{0.08}$, in agreement with experimental data.
Finally, the quantity $C_E$ takes into account the modification of the eikonal due to inelastic diffractive states. 
In refs. \cite{6r} and \cite{16r} a value $C_E = 1.5$ is used corresponding to a 50\% contribution of low-mass diffractive states relative to the elastic ones. 
Note that here we have also included in $C_E$ high-mass diffraction and taken $C_E = 1.8$.
Actually, the percentage contribution of large-mass diffractive states, given by the triple Pomeron graph, increases with energy. However, this energy dependence turns out to be small in the LHC energy range\footnote{Integrating the $M^2$ dependence of the triple Pomeron $[M^2]^{-1-\Delta}$ from $M^2 = 5$~GeV$^2$ to $s/20$, with $\Delta = 0.19$ we get a 6\% increase between $\sqrt{s} = 2.76$ and 14 TeV.}. Note also that including the large-mass diffraction contribution in $C_E$ is only possible if the $t$-slopes of the two-Pomeron-exchange graph and the triple Pomeron graph are equal\footnote{It should also be mentioned that with $\Delta \not= 0$ the signature factor induces a well defined real part in the scattering amplitude. However, only its imaginary part contributes to the non-diffractive observables considered in this paper.}.
Diffractive cuts corresponding to $k=0$ have not been included. A treatment of diffractiction would require a 
more refined approach, proceeding
for instance along the lines of refs. \cite{13r,Kai2010}.\par 

\begin{figure*}[htb!]
\vskip -5.25cm
\begin{center}
\begin{flushleft}
\includegraphics[width=1.\textwidth]{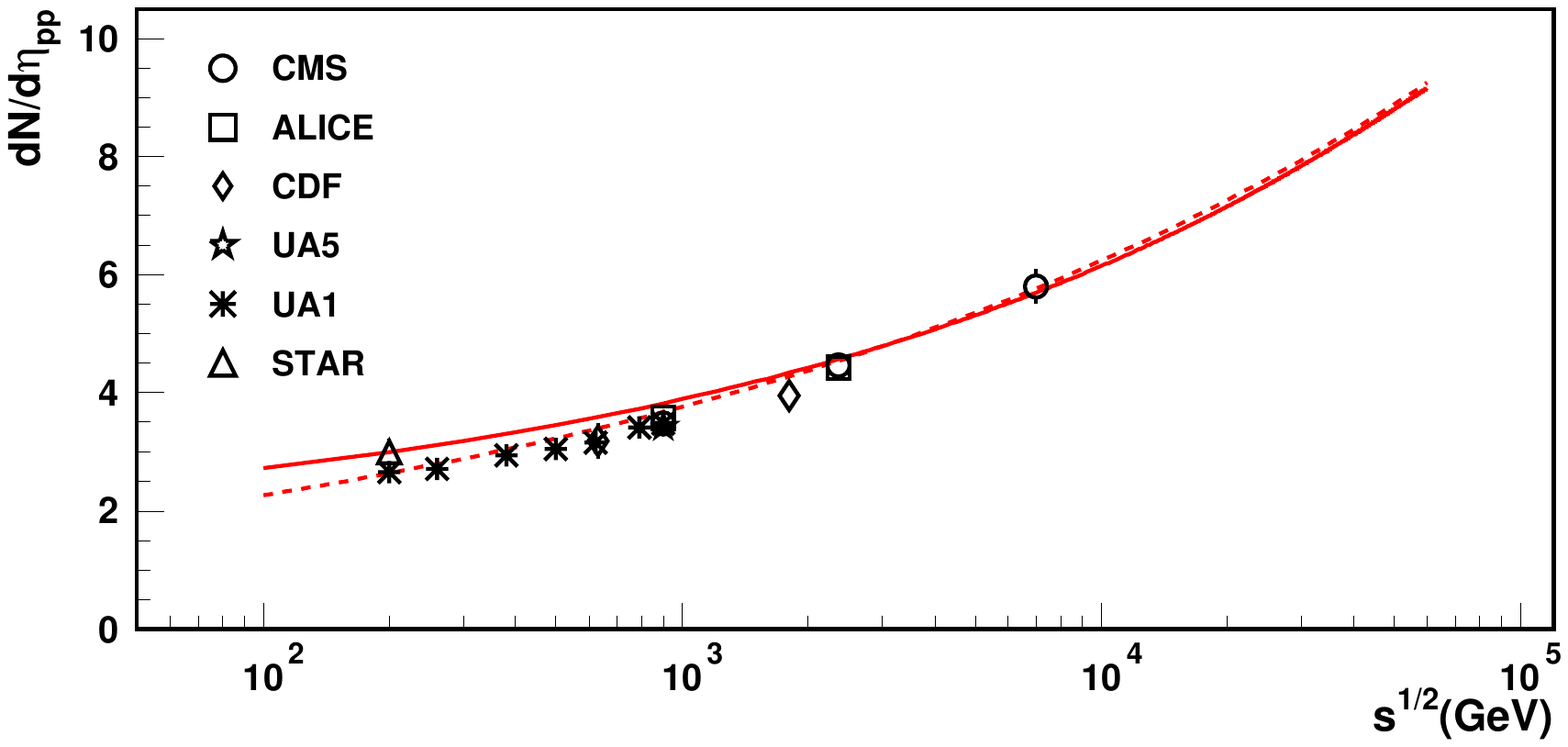}
\end{flushleft}
\end{center}
\vskip -12cm
\caption{ The values of $dN^{pp}/d\eta$ 
obtained from the formulae in Section 3 are plotted as a function of $\sqrt{s}$ in the range $\sqrt{s} = 100$ GeV 
to 50 TeV and compared to non-single-diffractive (NSD) data \cite{1r,3r,CDF,UA5,UA1,STAR}.
The line $s^{0.11}$ (dashed line)
representing the available experimental data 
is also plotted.}
\label{fig3}
\vskip -0.5cm
\end{figure*}

In string models the proportionality between the $pp$ multiplicities and $\overline{\kappa}$ mentioned above is expressed as \cite{5r} $dN^{pp}/d\eta = 2 \overline{\kappa} dN_S/d\eta$, where the last factor is the multiplicity in a single string. This relation is valid at high energies --when all strings have equal mid-rapidity multiplicity. $dN_S/d\eta (y^* = 0)$ is a constant, independent of $s$. In the calculations we have used $2dN_S/d\eta = 1.5$\footnote{This is to be compared with a value of  $2dN_S/dy (y^* = 0)$ about 2 used in DPM, corresponding to a plateau height of a single string close to one \cite{7r}. Note that the ratio $dN^{pp}/dy$ over $dN^{pp}/d\eta$ is close to 1.3 at $y^* = 0$.} obtained from the experimental value $dN^{pp}/d\eta = 4.7$ at $\sqrt{s} = 2.76$~TeV \cite{1r}. 
The results for $\sigma_{pp}^{ND}$ and 
$dN^{pp}/d\eta (y^*=0)$ as a function of $s$ are given in Table~\ref{tab1} and Fig.~\ref{fig3}. 
\begin{table}[htb!]
\begin{center}\setlength{\arrayrulewidth}{1pt}
\caption{The values of the non-diffractive $pp$ cross-section and charged particle pseudo-rapidity densities in the central rapidity region for the energy range between 200 and 50000 GeV.}
\label{tab1}
\vskip 0.75cm
\begin{tabular}{ccccccc}
\hline\hline
 $\sqrt{s}$ (GeV) & $dN^{pp}/d\eta (y^*=0)$ & $\sigma_{pp}^{ND}$ (mb)\\
\hline
200 &  2.99 & 31.22 \\
540 & 3.50 & 38.97 \\
900 & 3.82 & 43.33 \\
1800 & 4.34 & 49.64 \\
2760 & 4.71 & 53.77 \\
5500 & 5.42 & 60.78 \\
7000 & 5.70 & 63.33 \\
14000 & 6.61 & 70.99 \\
50000 & 8.80 & 86.19 \\
\hline\hline
\end{tabular}
\end{center}
\end{table}

We see that in the LHC energy region and beyond the $pp$ multiplicity follows closely the $s^{0.11}$ behaviour observed in the data \cite{1r}. Actually, this energy behaviour is predicted to hold in a much larger energy range --at least up to 50 TeV. As a consequence $\sigma_{pp}^{ND}$ follows closely an $s^{0.08}$ energy dependence. Note that at lower energies the calculated multiplicity is larger than the measured one. This is consistent with energy-momentum conservation effects which reduce the former one. At $\sqrt{s} = 200$~GeV the difference between calculated and observed multiplicity is about 15\%. 

Note that our calculated values of $dN^{pp}/dy$ refer to non-diffractive (ND) multiplicities. However, 
the available experimental data on multiplicities refer
to non-single-diffractive (NSD) interactions, or to inelastic (INEL) ones.
Taking into account the smallness of the double-diffractive contribution (DD),
\cite{1r,3r,UA5sigma,CDFsigma},
the difference bewteen non-diffractive (ND) and non-single-diffractive (NSD) multiplicities will be 
small.
Therefore, 
we have compared our theoretical non-diffractive value (ND) to experimental non-single-diffractive (NSD) data.
\par

\section{Multiplicities in PbPb collisions}
\hspace*{\parindent} 
The AGK cancellation implies that at mid-rapidities $d\sigma^{AA}/d\eta = A^2d\sigma^{pp}/d\eta$, which implies
\beq
\label{6e}
{dN^{AA} \over d\eta} = {A^2\ \frac{\sigma_{pp}^{ND}}{\sigma_{AA}}}\ {dN^{pp} \over d\eta} = n_{coll} \ {dN^{pp} \over d\eta} 
\eeq

\noi where $n_{coll} = A^2\ \sigma^{ND}_{pp}/ \sigma_{AA}$ is the average number of binary nucleon-nucleon collisions. As a function of impact parameter we have 
\beq
\label{7e}
n_{coll}(b) = A^2\ {\sigma^{ND}_{pp} \over \sigma_{AA}(b)}\ T_{AA}(b)
\eeq

\noi with $\sigma_{AA}(b) = 1 - \exp [- \sigma^{ND}_{pp} A^2 \ T_{AA}(b)]$ and $T_{AA}(b) = \int d^2s \ T_A (s) \ T_A(b-s)$ (see Section 2). It follows from eqs. (\ref{6e}) and (\ref{7e}) that the multiplicity per participant pair is given by
\beq
\label{8e}
R(b) = {dN^{AA} \over d\eta}(b) /(n_{part} /2) = {2n_{coll}(b) \over n_{part}(b)}\ {dN^{pp} \over d\eta}
\eeq 

\noi As discussed in the Introduction, the first factor in eq. (\ref{8e}) is about 5 at RHIC for central AuAu collisions and increases with $s$. This is much larger than the factor 2 observed experimentally. However, as discussed in Section 2, the AGK cancellation is only valid in the absence of shadowing. In its presence, eq. (\ref{6e}) has to be modified. One has
\beq
\label{9e}
{dN^{AA} \over d\eta}(b) = {A^2 \frac{\sigma_{pp}^{ND}}{\sigma_{AA}(b)}}\ {dN^{pp} \over d\eta} \ S^{sh}(b)
\eeq

\noi where the shadowing correction $S^{sh}(b)$ is given by eqs. (\ref{3e}) and (\ref{4e}). In this way, the $AA$ multiplicity is strongly reduced \cite{7r, 9r}. This reduction is given by the suppression factor $S^{sh}(b)/T_{AA}(b)$. Its numerical values for PbPb collisions at $\sqrt{s} = 2.76$ and 5.5 TeV are given in Table \ref{tab2} as a function of $b$. 
\begin{figure*}[htb!]
\begin{center}
\includegraphics[width=0.75\textwidth]{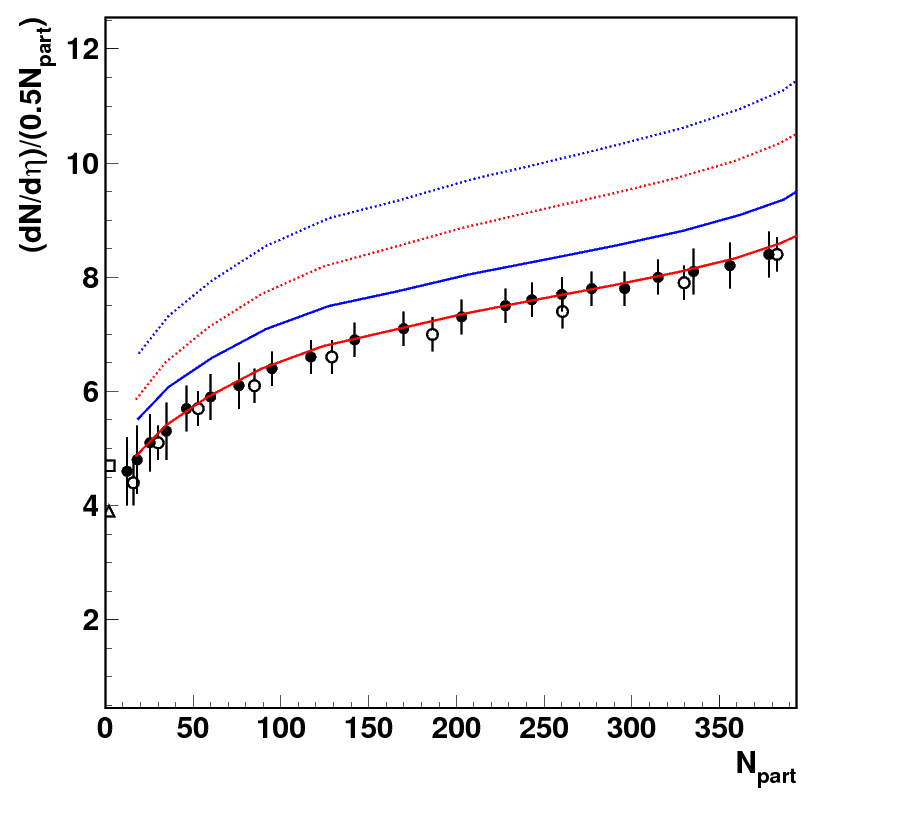}
\end{center}
\caption{The values of $dN^{AA}/d\eta$ at $\sqrt{s} = 2.76$ and 5.5 TeV are plotted for different centrality bins. The dashed lines are obtained from eq.~(\ref{9e}) with shadowing corrections given by eqs. (\ref{3e}) and(\ref{4e}) and using the non-diffractive values $dN^{pp}/d\eta = 4.7$ at $\sqrt{s} = 2.76$ GeV and 5.4 at $\sqrt{s} = 5.5$ TeV. The full lines are obtained using instead the inelastic values $dN^{pp}/d\eta = 3.9$ at $\sqrt{s} = 2.76$ GeV and 4.5 at $\sqrt{s} = 5.5$ TeV (see main text). Data are from \cite{Aamodt:2010cz} (open symbols) and \cite{:2011yr} (full symbols). The values for inelastic and non-single-diffractive $dN^{pp}/d\eta$ at $\sqrt{s} = 2.76$ TeV are also shown \cite{2r}.}
\label{fig4}
\end{figure*}
\begin{table}[htb!]
\begin{center}\setlength{\arrayrulewidth}{1pt}
\caption{The values of the shadowing suppression factor $S^{sh}(b)/T_{AA} (b)$ for PbPb collisions at $\sqrt{s} = 2.76$ and 5.5 TeV, as a function of the impact parameter $b$, computed from eqs.~(\ref{3e}) and (\ref{4e}).}
\label{tab2}
\vskip 0.75cm
\begin{tabular}{ccccccc}
\hline\hline
b & $\sqrt{s}=2.76$ TeV & $\sqrt{s}=5.5$ TeV \\
\hline
0 &  0.2828    &    0.2390 \\ 
2 &  0.2839    &    0.2396 \\
4 &  0.2910    &    0.2467 \\
6 &  0.3091    &    0.2637 \\
8 &  0.3432    &    0.2959 \\
10 &  0.4091   &    0.3599 \\ 
13 &  0.6167   &    0.5714 \\
\hline\hline
\end{tabular}
\end{center}
\end{table}

Using the numerical values in Tables~\ref{tab1} and \ref{tab2} we can compute the mid-rapidity $AA$ multiplicities for any value of the impact parameter.  The results at $\sqrt{s} = 2.76$ and 5.5 TeV are shown in Fig.~4 (dashed lines). 
At $\sqrt{s} = 2.76$~TeV the centrality dependence of the data \cite{2r} is well reproduced. Practically the same centrality dependence is predicted at $\sqrt{s} = 5.5$~TeV. The same feature has been observed experimentally between $\sqrt{s} = 200$~GeV and 2.76 TeV \cite{2r}. Its increase between $\sqrt{s}=2.76$ and 5.5 TeV is slightly less than the corresponding increase in the $pp$ multiplicity. An increase larger than the one in $pp$ has been observed between $\sqrt{s} = 200$~GeV and 2.76 TeV.\par

The absolute value of the multiplicity at $\sqrt{s} = 2.76$~TeV is about 25\% higher than the experimental one. Note, however, that there is an uncertainty in this absolute value due mainly to the uncertainty in $\sigma_{pp}^{ND}$. Moreover, there is a puzzling feature in the data \cite{Aamodt:2010cz}, namely the ratio between the two extreme centrality bins multiplicities is smaller than the one between the most central bin and $pp$. Therefore the multiplicity in peripheral bins gets smaller than the $pp$ one. Actually the data extrapolate nicely to the inelastic $pp$ multiplicity (3.9 at $\sqrt{s} = 2.76$~TeV) rather than to the non-diffractive one (4.7 at $\sqrt{s} = 2.76$~TeV). Using the former value in our calculations, instead of the non-diffractive one, we get absolute values in agreement with experiment.

\section{Conclusions}
\hspace*{\parindent} 
We have studied the mid-rapidity charged particle multiplicities in $pp$ and $AA$ collisions in the LHC energy range. We have shown that the observed energy dependence of the $pp$ multiplicity requires a Pomeron intercept close to 1.2. One obtains in this way an energy dependence remarkably close to $s^{0.11}$, in good agreement with available data. This behaviour is predicted to hold up to at least 50 TeV. At lower energies the calculated $pp$ multiplicity is larger than the experimental one. The difference is about 15~\% at $\sqrt{s} = 200$~GeV. This behaviour is consistent with energy conservation effects which reduce the multiplicity and are present at low energies. These effects go away when we reach the LHC range.\par 

In $AA$ collisions the centrality dependence of the multiplicity is predicted to be practically identical at $\sqrt{s} = 2.76$ and 5.5 TeV, 
in agreement with lower energy extrapolations. Its size is well reproduced. 
The increase in multiplicity with energy between 2.76 and 5.5 TeV is predicted to be slightly less than the one in $pp$,
while a larger increase is expected from lower energy results which show an energy dependence in $s^{0.15}$ for $AA$ and $s^{0.11}$ for $pp$ \cite{Aamodt:2010cz}. \par

The obtained Pomeron intercept is larger than the conventional one used in most models obtained from lower energy data. In this respect one can consider two different scenarios. A first possibility \cite{Kai2} is that a Pomeron intercept close to 1.2 can describe the data at all energies when all multiple scattering contributions are taken into account --together with low energy corrections such as energy conservation effects in the multiplicities. A second scenario is the so-called Pomeron flavoring \cite{18r}. In this scenario the Pomeron intercept increases due to the opening up of effective thresholds for heavy flavor production as the energy increases. Actually the incorporation of semi-hard events (mini-jets) within DPM \cite{19r,20r} can be considered a flavoring phenomenon \cite{21r}. \par


Calculations of the $pp$ multiplicity at LHC in the framework of the QGSM \cite{6r} can
be found in \cite{22r}. In these calculations the Pomeron intercept is taken to be 1.12.
As a consequence, the energy dependence of the $pp$ multiplicity is
substantially smaller than ours.
Monte Carlo models such as PYTHIA \cite{Pythia}
and PHOJET \cite{Phojet} also have a too small energy dependence in the LHC range \cite{3r}. 
All these results lend support to the necessity of a higher Pomeron
intercept.
Obviously our value of an intercept close to 1.2 has to be validated by a measurement of the energy dependence up to 14 TeV.

\vskip 1cm
\section*{Acknowledgments}
\hspace*{\parindent} 
We dedicate this paper to the memory of our friend and long-time collaborator Alexei Kaidalov. He was a world expert in the field discussed in this paper, to which he has made outstanding contributions. We have benefited from his creativity and deep knowledge of the field during many years of fruitful and stimulating collaboration.

This work is partially supported by MEC/IN2P3 
(AIC10-D-000569 and AIC-D-2011-0740). E. G. F. thanks Javier L. Albacete for fruitful and supporting discussions during the writing of this paper.

\end{document}